\documentclass{appolb}
\usepackage{graphicx}
\usepackage[latin1]{inputenc}
\usepackage{amsmath}
\usepackage{amsfonts}
\usepackage{amssymb}
\usepackage{graphicx}
\usepackage{cite}
\usepackage{bm}
\usepackage{slashed}
\usepackage{hyperref}
\usepackage{float}
\usepackage{placeins}
\usepackage{flafter}
\usepackage{caption}
\usepackage{subcaption}
\usepackage{lineno,hyperref}
\usepackage{longtable}
\usepackage{enumerate}
\graphicspath{{figures/}}

\begin{document}
\title{Predictions of $\alpha$-decay half-lives for neutron-deficient nuclei with the aid of artificial neural network%
}
\author{A. A. Saeed, W. A. Yahya\footnote{email: wasiu.yahya@gmail.com}, O. K. Azeez
\address{$^\dagger$Department of Physics and Materials Science, Kwara State University, Malete, Nigeria
} 
}
\maketitle
\begin{abstract}
In recent years, artificial neural network (ANN) has been successfully applied in nuclear physics and some other areas of physics. This study begins with the calculations of $\alpha$-decay half-lives for some neutron-deficient nuclei using Coulomb and proximity potential model (CPPM), temperature dependent Coulomb and proximity potential model (CPPMT),  Royer empirical formula, new Ren B (NRB) formula, and a trained artificial neural network model ($T^{ANN}$). By comparison with experimental values, the ANN model is found to give very good descriptions of the half-lives of the neutron-deficient nuclei. Moreover CPPMT is found to perform better than CPPM, indicating the importance of employing temperature-dependent nuclear potential. Furthermore, to predict the $\alpha$-decay half-lives of unmeasured neutron-deficient nuclei, another ANN algorithm is trained to predict the $Q_{\alpha}$ values. The results of the $Q_{\alpha}$ predictions are compared with the Weizs$\ddot{a}$cker-Skyrme-4+RBF (WS4+RBF) formula. The half-lives of unmeasured neutron-deficient nuclei are then predicted using CPPM, CPPMT, Royer, NRB, and $T^{ANN}$, with $Q_{\alpha}$ values predicted by ANN as inputs. This study concludes that half-lives of $\alpha$-decay from neutron-deficient nuclei can successfully be predicted using ANN, and this can contribute to the determination of nuclei at the driplines.

\end{abstract}
  
\section{Introduction}

$\alpha$-decay is one of the most important types of radioactive decay in the study of nuclei \cite{Santhosh2020d}, owing to its ability to provide insights on nuclear structure and stability of nuclei \cite{Cheng2019}.  It was discovered by Ernest Rutherford in 1899 as a component out of three components of radiation emitted by uranium nucleus \cite{Zdeb2013}.  In 1928,  Gamow \cite{GAMOW1928}, Gurney and Condon \cite{gurney1928, gurney1929} gave a theoretical explanation of the Geiger-Nuttall law,  which derives its basis from the quantum tunneling effect and was the first successful attempt at the quantum description of nuclear phenomena. Since then,  many theoretical models and empirical formulas have been proposed to calculate $\alpha$-decay half-lives of nuclei. Some of the theoretical models include the effective liquid drop model (ELDM) \cite{Tavares1998,Cui2021},  the generalized liquid drop model (GLDM) \cite{royer1985,Bao2014,NaNa2016}, modified generalized liquid drop model (MGLM) \cite{kp2019,Dashty2019}, preformed cluster model (PCM) \cite{GUPTA1994,Singh2010}, fission-like model \cite{fei2010}, and so on. Some of the theoretical models use phenomenological potentials while others use microscopic potentials \cite{Yahya2021,Yahyapol2021}.
Some of the empirical formulas that have been successful in the investigation of $\alpha$-decay half-lives are the Royer formula \cite{Royer_2000,Royer2010,deng2020}, Denisov and Khudenko formula \cite{DK2010},  Viola and Seaborg formula (VSS) \cite{viola1966}, Ren formulas \cite{Ren2004} (and the modified Ren formulas \cite{Akrawy2019}), universal decay law (UDL) \cite{Qi2009}, Akrawy and Poenaru formula \cite{Akrawy2017}, etc.\\

$\alpha$-decay is a dominant radioactive decay mode for unstable nuclei, particularly neutron-deficient nuclei.  An atomic nucleus is said to be neutron-deficient if it consists of more protons than neutrons; they are also called proton-rich nuclei, and are close to the proton drip-line. Most of the observed neutron-deficient nuclei with mass number $A \geq 150$ can undergo $\alpha$-decay. The study of the half-lives of neutron-deficient nuclei can contribute to the determination of nuclei at the driplines. The contribution to the determination of nuclei at the driplines have motivated some researches on nuclei with $Z > N$ \cite{wangetal2017,Cui2019,chong2016,Gao2020,Ahmed2021}. This study will calculate the $\alpha$-decay half-lives of neutron-deficient nuclei using Coulomb and proximity potential model and two empirical formulas. It is known that the Coulomb and proximity potential model (CPPM), and temperature dependent Coulomb and proximity potential model (CPPMT) are successful models in the investigation of $\alpha$-decay half-lives \cite{Manjunatha2015,WAYahya2020,Zanganah2020,Santhosh2012}. The two empirical formulas to be employed are the Royer formula \cite{Royer_2000,Royer2010} and new Ren B formula \cite{Akrawy2019}. These formulas have been known to be successful in the calculation of $\alpha$-decay half-lives of nuclei.  \\

In recent years,  machine learning has grown in popularity in the physics community due to its ability to learn from data and arrive at reasonable conclusions.  The two commonly used techniques in machine learning are supervised and unsupervised learning techniques.  In supervised learning, data with labels are used to train the model with the goal of predicting outcomes as accurately as possible. In unsupervised learning, data with no labels are fed into the model with the goal of finding hidden patterns in the data and arriving at reasonable conclusions.  Artificial neural network (ANN), which is an algorithm under supervised machine learning contains a large system that is created and programmed to mimic the human brain \cite{Leonardo2019}, and operates by using dense layers made up of neurons to process information. These neurons are also known as units and are arranged in series. The data go through the input layer of the ANN from external sources for the system to learn from and this information is processed in the hidden layer connected by weights, which then becomes the outcome in the output layer.  There have been some successful applications of machine learning in nuclear physics. For example machine learning and deep learning have been employed in the study of nuclear charge radii \cite{Akkoyun_2013,Wu2020}, in the predictions of nuclear $\beta$-decay half-lives \cite{Niu2019}, in the extraction of electron scattering cross-sections from swarm data \cite{Jetly2021}, in the shell model calculations for proton-rich zinc (Zn) isotopes \cite{Serkan2019}, and in the prediction of $\alpha$-decay $Q_\alpha$ values \cite{Ubaldo2019}. \\

In this paper, we employ the use of the Coulomb and proximity potential model (CPPM) in the calculation of the $\alpha$-decay half-lives of some measured neutron-deficient nuclei. Since it is known that the use of temperature-dependent potential can improve the prediction of $\alpha$-decay half-lives, we have also used the temperature-dependent Coulomb and proximity potential model (termed CPPMT). Moreover, an artificial neural network (specifically, a multilayer feed forward neural network) is also used to predict the half-lives. Two empirical formulas viz. Royer and new Ren B formulas have also been used to determine the performance accuracy of the CPPM, CPPMT, and ANN models. Since the NUBASE2020 \cite{Kondev_2021,Wang2021} database is now publicly available, the data used in the study have been extracted from the database. New coefficients for the two empirical formulas are determined using a least square fit scheme with input data from the NUBASE2020 database. The study also predicts the half-lives of $\alpha$-decay from some unmeasured neutron-deficient nuclei. To achieve this, one requires the $Q_{\alpha}$ values for the $\alpha$-decay processes. Since there are no experimental $Q_{\alpha}$ values for the unmeasured neutron-deficient nuclei, an artificial neutral network (ANN) has been trained to predict $Q_{\alpha}$ values using about $1021$ $Q_{\alpha}$ values from the NUBASE2020 database. The trained ANN model was then used to predict the $Q_{\alpha}$ values for unmeasured neutron-deficient nuclei. The results obtained are compared to the theoretical WS4 and WS4+RBF \cite{Wang2014} $Q_{\alpha}$ values. The predicted $Q_{\alpha}$ values are then used as inputs to predict the $\alpha$-decay half-lives of unmeasured neutron-deficient nuclei. \\

The paper is presented as follows:  the theoretical models used are introduced in Section \ref{theory}. In Section \ref{results}, the results of the calculations are presented and discussed, and in Section \ref{conclusion}, the conclusion is presented. 

\section{Theoretical Formalism}
\label{theory}
\subsection{Coulomb and proximity potential model (CPPM)}
In this model, the total interaction potential between the $\alpha$ particle and the daughter nucleus can be expressed as the summation of the proximity potential, Coulomb potential, and centrifugal potential for both the touching configuration and separated fragments. That is \cite{Manjunatha2015}:
\begin{equation}
V = V_C(r) + V_P(z) + \frac{\hbar \ell \left(\ell + 1 \right)}{2\mu r^2} ,
\end{equation}
where the last term is the centrifugal potential, $\ell$ is the angular momentum carried by the $\alpha$ particle, the Coulomb potential $V_C$ given by: 
\begin{equation}
V_C(r) = Z_1 Z_2 e^2 \begin{cases}
\frac{1}{r} & \text{for} \ r \ge R_C\\
\frac{1}{2R_C}\left[3 - (\frac{r}{R_C})^2\right] & \text{for} \ r \le R_C\\ 
\end{cases}.
\end{equation}
Here, $Z_1$ and $Z_2$ represent the charge number of the $\alpha$ particle emitted and the daughter nucleus, respectively, and $r$ is the distance between the fragments centres.  $R_C$ is known as the radial distance and is given by $R_C = 1.24(R_1 + R_2)$, where $R_1$ and $R_2$ are defined below.  \\

The first recorded implementation of the proximity potential was in 1987 by Shi and Swiatecki where nuclear deformation influence, and the shell effects on the half-life of exotic radioactivity were estimated \cite{Shi1987}. Two years later,  Malik et al. applied the proximity potential model in the preformed cluster model \cite{Malik1989}. The calculation of the strength of the interaction of the daughter and emitted $\alpha$ particle yields the proximity potential $V_P(z)$ provided by Blocki et al \cite{Bocki1977},  and is given as:
\begin{equation}
V_P(z) = 4\pi\gamma b \bar{R} \phi \left(\frac{z}{b}\right) \mathrm{MeV},
\label{2.3}
\end{equation}
where the nuclear surface potential $\gamma$ is given as
\begin{equation}
\gamma = 1.460734\left[1 - 4\left(\frac{N - Z}{N + Z}\right)^2\right] \mathrm{MeV/fm}^2.
\end{equation}
Here $N$ and $Z$ denote the neutron number and proton number of the parent nucleus, respectively.  $\phi$ is the universal proximity potential, given by \cite{Gharaei2016}: 
\begin{equation}
\phi (\epsilon) = \begin{cases}
\frac{1}{2} (\epsilon - 2.54)^2 - 0.0852(\epsilon - 2.54)^3 & \ \epsilon \le 1.2511\\
-3.437 \exp(-\epsilon/0.75) & \ \epsilon \le 1.2511\\
\end{cases},
\label{2.5}
\end{equation}
where $\bar{R}$ is called the mean curvature radius, and it is dependent on the form of both nuclei.  It can be expressed as: 
\begin{equation}
\bar{R} = \frac{C_1C_2}{C_1 + C_2},
\end{equation}
$C_1$ and $C_2$, known as the Süsmann central radii, are calculated using
\begin{equation}
C_i = R_i - \left(\frac{b^2}{R_i}\right),
\label{2.7}
\end{equation}
where $R_i$ can be obtained with the aid of a semi-empirical formula in terms of mass number $A_i$ \cite{Bocki1977}: 
\begin{equation}
R_i  = 1.28A_{i}^{1/3} - 0.76 + 0.8A_{i}^{-1/3}.
\end{equation}
The penetration probability of the $\alpha$ particle through the potential barrier can be determined with the aid of the WKB approximation \cite{Gurvitz1987,WAYahya2020}:
\begin{equation}
P = \exp \left[-\frac{2}{\hbar}\int \limits_{R_i}^{R_o} \sqrt{2\mu[V(r) - Q]dr}\right],
\label{2.10}
\end{equation}
where $\mu = A_1 A_2 /A$ is the reduced mass, $A_1$ and $A_2$ are the mass numbers of emitted $\alpha$ particle and daughter nucleus, respectively, $A$ is the mass number of the parent nucleus, $R_i$ and $R_o$ are known as the classic turning points, obtained via:
\begin{equation}
V(R_i) = V(R_o) = Q.
\label{2.11}
\end{equation}

The $\alpha$-decay half-life can finally be calculated via: 
\begin{equation}
T_{1/2} = \frac{\ln 2}{\lambda},
\end{equation}
where $\lambda = \nu P $, and $\nu = 10^{20} s^{-1}$ is known as the assault frequency.

\subsection{Temperature dependent Coulomb and proximity potential model (CPPMT)}
The temperature dependent proximity potentials can be written as:
\begin{equation}
V_P(r, T) = 4 \pi \gamma (T) b(T) \bar{R} (T) \phi (\xi).
\label{proxT}
\end{equation}
Here $\phi (\xi)$ is still the universal function, the temperature dependent forms of the other parameters in equation (\ref{proxT}) are given by \cite{Salehi2013,Sauer1976,Shlomo1991,WAYahya2020}:
\begin{gather}
\gamma (T) = \gamma(0) \left(1 - \frac{T - T_b}{T_b}\right)^{3/2},\\
b(T) = b(0)(1 + 0.009T^2),\\
R(T) = R(0) (1 + 0.0005 T^2).
\end{gather}
Here $T_b$ is the temperature that is associated with near Coulomb barrier energies.  A different version of the temperature dependent surface energy coefficient given by $\gamma (T) = \gamma (0)(1 - 0.07T)^2 $ \cite{WAYahya2020} has been used in this work. The temperature T (MeV) can be derived from: \cite{Gupta_1992,Puri_1992}
\begin{equation}
E^{*} = E_{kin} + Q_{in} = \frac{1}{9}AT^2 - T,
\label{2.19}
\end{equation}
where $E^{*}$ denotes the parent nucleus excitation energy, and $A$ is its mass number. $Q_{in}$ represents the entrance channel Q-value of the system.  $E_{kin}$ is the kinetic energy of the $\alpha$ particle emitted and can be obtained using \cite{WAYahya2020}:
\begin{equation}
E_{kin} = (A_2/A)Q.
\label{2.20}
\end{equation}

\subsection{Royer empirical formula}
In the year 2000, Royer \cite{Royer_2000} proposed an analytical formula for the calculation of the $\alpha$-decay half-lives of nuclei, by applying a fitting procedure to some $\alpha$ emitters. The proposed formula did not contain dependence on the angular momentum carried by the $\alpha$ particle. In the year 2010, Royer proposed an improved formula for calculating the $\alpha$-decay half-lives, which is explicitly dependent on the angular momentum ($\ell$) carried by the $\alpha$ particle. The angular momentum for even-even nuclei was taken to be zero. It was observed that the agreement with experimental data was better than what was earlier recorded. The proposed formula is given for even-even, even-odd, odd-even, and odd-odd nuclei as \cite{Royer2010}:\\
\begin{equation}
\log_{10}[T] = -25.752 - 1.15055A^{\frac{1}{6}} \sqrt{Z} + \frac{1.5913Z}{\sqrt{Q}},
\label{royer1}
\end{equation}
\begin{align}
\log_{10}[T] &= -27.750 - 1.1138A^{\frac{1}{6}} \sqrt{Z} + \frac{1.6378Z}{\sqrt{Q}} \\ \nonumber	
& + \frac{1.7383 \times 10^{-6} ANZ[\ell(\ell + 1)]^{\frac{1}{4}}}{Q} + 0.002457 A[1 - (-1)^\ell],
\label{royer2}
\end{align}
\begin{align}
\log_{10}[T] &= -27.915 - 1.1292A^{\frac{1}{6}} \sqrt{Z} + \frac{1.6531Z}{\sqrt{Q}} \\	\nonumber 
&+ \frac{8.9785 \times 10^{-7} ANZ[\ell(\ell + 1)]^{\frac{1}{4}}}{Q} + 0.002513 A[1 - (-1)^\ell],
\label{royer3}
\end{align}
\begin{align}
\log_{10}[T] &= -26.448 - 1.1023A^{\frac{1}{6}} \sqrt{Z} + \frac{1.5967Z}{\sqrt{Q}} \\  \nonumber
&+ \frac{1.6961 \times 10^{-6} ANZ[\ell(\ell + 1)]^{\frac{1}{4}}}{Q} + 0.00101 A[1 - (-1)^\ell] ,
\label{royer4}
\end{align}
respectively. The short form of equations (\ref{royer1}) -- (\ref{royer3}) can be written as:
\begin{equation}
\log_{10}[T] = a + b A^{\frac{1}{6}} \sqrt{Z} + \frac{c Z}{\sqrt{Q}} + \frac{d \times 10^{-6} A N Z [\ell(\ell + 1)]^{\frac{1}{4}}}{Q} + e A \left[ 1 - (-1)^\ell \right] ,
\label{royer5}
\end{equation}
where $a$, $b$, $c$, $d$, $e$ are the coefficients given in equations (\ref{royer1}) -- (\ref{royer3}) for even-even, even-odd, odd-even, and odd-odd nuclei, respectively. For even-even nuclei, $d = e = 0$.

\subsection{New Ren B (NRB) formula}
In 2018, Akrawy et al. \cite{Akrawy2019}  studied the influence of nuclear isospin and angular momentum on $\alpha$-decay half-lives.  The existing Ren B formula by \cite{Ni2008}, were improved by including asymmetry and angular momentum terms.  With the aid of least square fit and experimental values of 365 nuclei, the authors obtained new coefficients for the Ren B formula. The New Ren B formula yielded better results in the calculation of $\alpha$-decay half-lives than the existing Ren B formula, when compared with the experimental data \cite{Akrawy2019}. The New Ren B formula is given as:
\begin{equation}
\log_{10} T_{1/2}^{NRB} = a \sqrt{\mu} Z_1 Z_2 Q^{-1/2} + b \sqrt{\mu Z_1 Z_2} + c + d I + e I^2 + f[\ell(\ell + 1)],
\end{equation}
where $ \mu $ is the reduced mass and the nuclear isospin asymmetry $ I = \frac{N - Z}{A}$. The two $\alpha$-decay empirical formulas used in this work are the Royer formula and the New Ren B formula.

\subsection{Artificial Neural Network (ANN)}
ANN is a multilayer neural network made up of an input layer, hidden layers, and an output layer. We label the structure of our ANN network as $[M_1,  M_2,  \cdots,  M_n]$, where $M_i$ is the number of neurons in the $i$th layer. $i = 1$ denotes the input layer while $i = n$ denotes the output layer. The outputs from the $i$th hidden layer are calculated using the formula:
\begin{equation}
h(\theta_{i}, X) = \mathrm{ReLU}(0.01 w^{(i)} h(\theta _{i-1}, X) + 0.01b^{(i)}),
\end{equation}
where $h(\theta _{i-1}, X)$ denotes the outputs from the previous layers, $w^{(i)}$ and $b^{(i)}$ represent the parameters of the network, and ReLU is the activation function used in the hidden layers. The ReLU  is an a non-linear function that helps improve the performance of the model. It has been chosen as the activation function for the hidden layers in this work, because of its ability to solve the problem of vanishing gradient. The outputs $h(\theta _{1}, X)$  of the input layer are basically the input data $X$. For a regression problem like in our case, activation functions are not required in the output layer, therefore the output of the ANN network can be expressed as:
\begin{equation}
y = g(\theta , X) =w^{(n)}h(\theta_{n-1}, X) + b^{(n)},
\end{equation}
where $\theta = \{w^{(1)}, b^{(1)}, \cdots, w^{(n)}, b^{(n)}\}$ represent the network parameters, $h(\theta_{n-1}, x)$ represent the outputs of the hidden layers, and $X$ denote the inputs. In this work, ANN models have been trained to predict both half-life and $Q_{\alpha}$ values for some neutron-deficient nuclei. For the ANN model trained to predict the half-lives, $M_1 = 4$ (consisting the mass number, charge number, orbital angular momentum, and $Q_{\alpha}$ values) and $M_n = 1$, and for the ANN model trained to predict $Q_{\alpha}$ values, $M_1 = 2$ (consisting the mass number and charge number of the nuclei) and $M_n = 1$. The output layer $M_n = 1$ because we are dealing with a regression problem.\\

It is important to observe  how well the ANN model performs during the training phase, a cost function is used to achieve this. The cost function evaluates the performance of the model by observing the difference between the predicted and actual values.  Learning takes place by reducing the cost function to the barest minimum, this is achieved with the aid of an optimizing algorithm.  Adam is one of the most widely used optimization algorithms. It has been employed in this work to derive the best values for the parameters in the ANN network, by modifying the parameters $w^{(i)}$ and $b^{(i)}$ for $i = 1, \cdots, n$ in the network until an acceptable value between the predicted and actual output is achieved. The root mean square error has been used as the cost function in this study. It can be expressed as:
\begin{equation}
\mathrm{RMSE(\theta)} = \sqrt{\frac{1}{N} \sum \limits_{i=1}^{N}\left[ Y_{i}^\mathrm{{expt}} - g(\theta,  X_i) \right]^2} ,
\label{equ25}
\end{equation}
where $Y_i^\mathrm{{expt}}$ denote the experimental values, $g(\theta,  X_i)$ are the predicted output values, and $N$ is the size of the training data set.\\

In training the ANN model to predict the half-lives, a total number $(N)$ of $549$ nuclei in the NUBASE2020 database have been used. As a result, a network structure of [4,50,100,50,1] has been chosen. During the training phase, the dataset was split into 80\% train set and 20 \% test set. The test set has been used to validate the performance of the trained model. The performance of the model can be improved, if necessary, by tweaking the parameters of the model before using it for predictions. \\

To predict the half-lives for unmeasured neutron-deficient nuclei, the $Q_{\alpha}$ values are required as part of the input data. These unmeasured neutron-deficient nuclei have no experimental $Q_\alpha$ values. The aid of machine learning is therefore sought to predict the $Q_\alpha$ values, which can subsequently be used to calculate the half-lives of the unmeasured neutron-deficient nuclei. In order to achieve this, an ANN model is trained using about 1021 $Q_\alpha$ values of measured nuclei in the NUBASE2020 database. The dataset is also split into 80 \% train set and 20 \% test set. As a result of the number of instances of the data, a network structure of [2,120,120,120,1] has been chosen, and the performance accuracy is also determined using root mean square error. \\

\section{Results and Discussion}
\label{results}
The results of the calculations of the $\alpha$-decay half-lives of some neutron-deficient nuclei are presented and discussed here. The calculations have been carried out using Coulomb and proximity potential model (CPPM), temperature dependent Coulomb and proximity potential model (CPPMT), Royer empirical formula (Royer), new Ren B emprical formula (NRB), and trained artificial neural network (ANN).\\

\begin{figure}[H]
	\centering
	\includegraphics[scale=0.7]{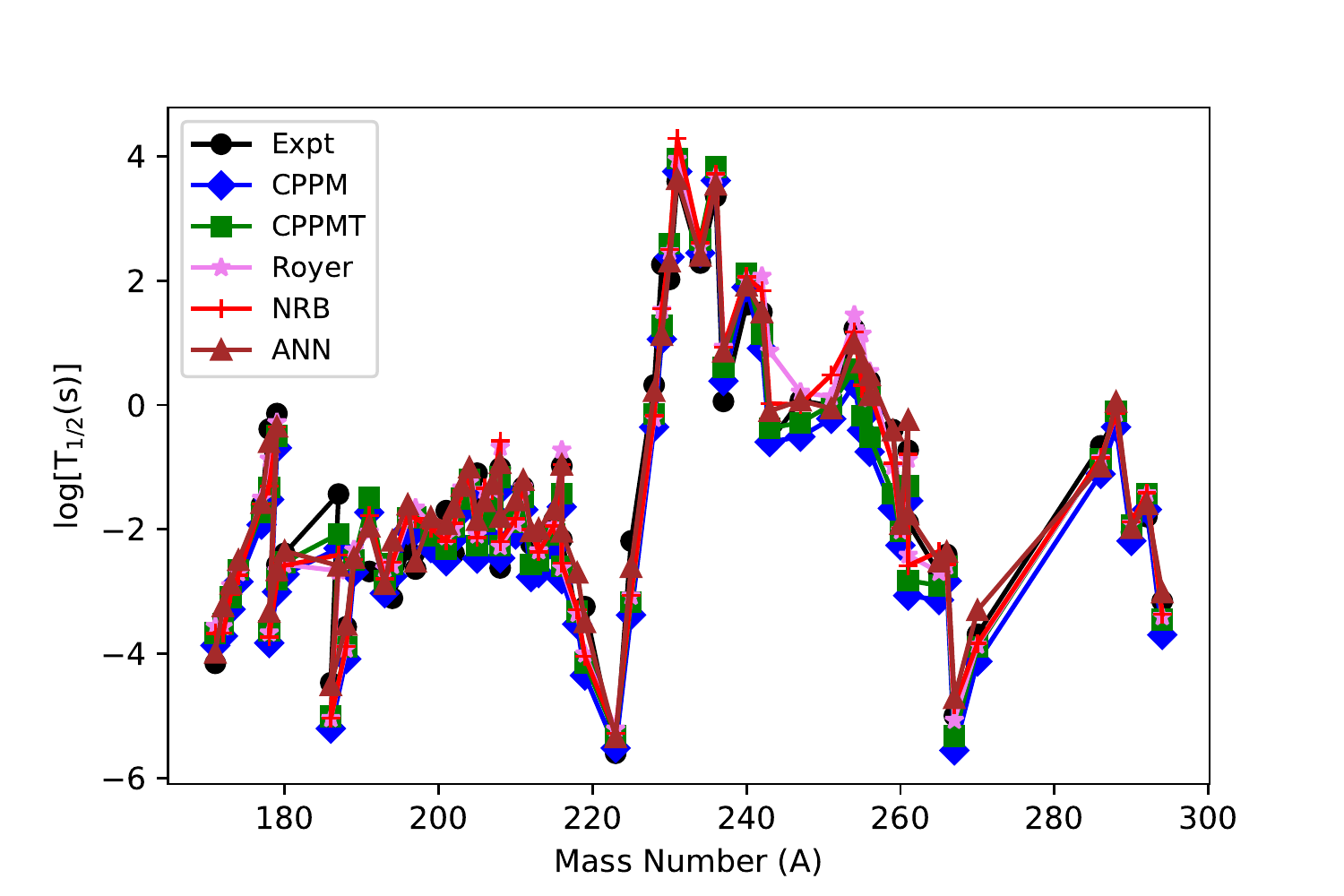} 
	\caption{ Plots of the experimental and theoretically calculated $\alpha$-decay half-lives for some neutron-deficient nuclei using CPPM, CPPMT, Royer, NRB, and ANN models.}
	\label{figure1}
\end{figure}

The coefficients given in Ref. \cite{Royer2010} for the Royer formula and Ref. \cite{Akrawy2019} for the new Ren B formula were obtained with the aid of a fitting procedure applied to the $\alpha$-decay half-lives in previous NUBASE databases. In this study, new coefficients have been obtained for the two formulas by applying the least square fit scheme and using $549$ $\alpha$ emitters in the NUBASE2020 database, containing $189$ even-even, $150$ even-odd, $117$ odd-even and $93$ odd-odd nuclei. The new coefficients obtained are given in Table \ref{table1} for Royer formula and Table \ref{table2} for new Ren B (NRB) formula. The root mean square error (RMSE) values obtained are $0.5411$ for Royer formula, and $0.5538$ for NRB formula.  \\
\begin{table}[H]
	\centering
	\caption{New coefficients for the Royer formula.}
	\begin{tabular}{cccccc}\hline
		\hline
		Nuclei &$a$ & $b$ & $c$ & $d$ & $e$\\ \hline
		\hline
		even-even & -25.5993 & -1.1362 &1.5771  & $0.0000 \times 10^0$ & $0.0000 \times 10^0$\\ 
		even-odd & -25.0031  & -1.1327 &  1.5622  &  $6.9116 {\times 10^{-7}}$  & $ 1.7000 \times 10^{-3} $ \\
		odd-even & -24.3063  & -1.1861 & 1.5787 & $9.7862{\times 10^{-7}}$  & $6.3120{\times 10^{-5}}$ \\
		odd-odd & -25.9529 & -1.1088 & 1.5848 & $9.7423 {\times 10^{-7}}$  & $-9.6530{\times 10^{-5}}$ \\ \hline
		\hline
	\end{tabular}
	\label{table1}
\end{table}
\begin{table}[H]
	\centering
	\caption{New coefficients for the new Ren B formula.}
	\begin{tabular}{ccccccc}\hline
		\hline
		Nuclei & $a$ & $b$ & $c$ & $d$ & $e$ & $f$\\ \hline
		\hline
		even-even & 0.4095 & -1.4111 & -15.2260  & 8.6687 & -49.5989 & 0.0000 \\ 
		even-odd & 0.4095  & - 1.3815 &  -14.8570  & -9.5116 & 27.5012 & 0.0323  \\
		odd-even & 0.4203  & -1.4093 & -15.3943 & -7.2587 & 07.1358 & 0.0298 \\
		odd-odd & 0.4135 & -1.4380 & -14.5336 & 1.7049 & 01.2728 & 0.0086 \\ \hline
		\hline
	\end{tabular}
	\label{table2}
\end{table}
To calculate the $\alpha$-decay half-lives for some neutron-deficient nuclei using ANN, the algorithm is trained using the data of $549$ $\alpha$ emitters. The train set contains $439$ nuclei while the test set contains $110$ nuclei. After training and optimizations, the values of the root mean square errors obtained for the train and test sets are shown in Table \ref{table3}.
\begin{table}[H]
	\centering
	\caption{The root mean square errors ($\sigma$) obtained for the training set and test set after training ANN to predict the half-lives.}
	\begin{tabular}{lr}\hline
		\hline
		Artificial Neural Network (ANN) & $\sigma$ \\ \hline
		\hline
		Train & 0.3876 \\ 
		Test & 0.5719\\
		\hline
	\end{tabular}
	\label{table3}
\end{table}
Having successfully trained the ANN model to predict $\alpha$-decay half-lives, the trained ANN model, CPPM, CPPMT, Royer, and NRB are now used to calculate the $\alpha$-decay half-lives of some neutron-deficient nuclei. Table \ref{table4} presents the results of the calculations. It can be observed that the values obtained from the five models are in good agreements with the experimental values.
\begin{longtable}{cccccccccccc}
	\caption{The Experimental and predicted log[T$_{1/2}$(s)] values for 69 neutron-deficient nuclei within the range of $80 \le Z \le 118$. }\\ \hline
	\hline
	& & & & \multicolumn{6}{c} {log$[T_{1/2} (s)]$} \\ \cline{5 - 10}
	A & Z & $Q_{\alpha}$&$\ell$ & Expt & CPPM & CPPMT & Royer & NRB & ANN \\
	\hline
	\hline
	\endfirsthead
	\multicolumn{9}{l}
	{\tablename\ \thetable\ -- \textit{Contd}} \\ \hline
	\hline
	\multicolumn{6}{r}{log$[T_{1/2} (s)]$} \\ \cline{5 - 10}
	A & Z & $Q_{\alpha}$&$\ell$ & Expt & CPPM & CPPMT & Royer & NRB & ANN  \\ \hline
	\hline
	\endhead
	\hline \multicolumn{10}{r}{\textit{Contd}} \\
	\endfoot
	\hline
	\endlastfoot
	\label{table4}
	
	171   & 80    & 7.6677 & 2     & -4.1549 & -3.8676 & -3.6663 & -3.5599 & -3.6724 & -3.9863 \\
	172   & 80    & 7.5238 & 0     & -3.6364 & -3.7166 & -3.5231 & -3.5659 & -3.6714 & -3.2377 \\
	173   & 80    & 7.3780 & 0     & -3.0969 & -3.2855 & -3.0928 & -2.9044 & -3.0393 & -2.8762 \\
	174   & 80    & 7.2333 & 0     & -2.6990 & -2.8430 & -2.6511 & -2.6975 & -2.7415 & -2.4969 \\
	177   & 81    & 7.0670 & 0     & -1.6081 & -1.9244 & -1.7315 & -1.4987 & -1.5908 & -1.5784 \\
	178   & 81    & 7.0200 & 2     & -0.3859 & -1.5200 & -1.3210 & -0.8672 & -1.3056 & -0.5979 \\
	179   & 81    & 6.7091 & 0     & -0.1377 & -0.6898 & -0.4987 & -0.2800 & -0.3497 & -0.3485 \\
	178   & 82    & 7.7895 & 0     & -3.6021 & -3.8289 & -3.6332 & -3.6638 & -3.7310 & -3.3282 \\
	179   & 82    & 7.5961 & 2     & -2.5686 & -3.0065 & -2.8046 & -2.6684 & -2.8173 & -2.6574 \\
	180   & 82    & 7.4187 & 0     & -2.3872 & -2.7246 & -2.5304 & -2.5653 & -2.5799 & -2.3492 \\
	187   & 83    & 7.7791 & 5     & -1.4318 & -2.3051 & -2.0748 & -2.6682 & -2.4127 & -2.5861 \\
	186   & 84    & 8.5012 & 0     & -4.4685 & -5.2045 & -5.0084 & -5.0413 & -5.0398 & -4.5078 \\
	188   & 84    & 8.0823 & 0     & -3.5686 & -4.0855 & -3.8900 & -3.9232 & -3.8862 & -3.5392 \\
	189   & 84    & 7.6943 & 2     & -2.4559 & -2.6921 & -2.4901 & -2.3272 & -2.4853 & -2.4598 \\
	191   & 85    & 7.8223 & 5     & -2.6778 & -1.7272 & -1.4932 & -2.0419 & -1.7761 & -1.9592 \\
	193   & 86    & 8.0400 & 2     & -2.9393 & -3.0276 & -2.8220 & -2.6313 & -2.7971 & -2.8824 \\
	194   & 86    & 7.8624 & 0     & -3.1079 & -2.7551 & -2.5566 & -2.5785 & -2.5265 & -2.1815 \\
	196   & 86    & 7.6167 & 0     & -2.3279 & -2.0125 & -1.8148 & -1.8480 & -1.7757 & -1.6018 \\
	197   & 87    & 7.8964 & 3     & -2.6383 & -2.0219 & -1.8081 & -1.6564 & -1.8255 & -2.5173 \\
	199   & 87    & 7.8168 & 0     & -2.1805 & -2.3000 & -2.1008 & -1.9129 & -1.9754 & -1.8068 \\
	201   & 88    & 8.0015 & 0     & -1.6990 & -2.5137 & -2.3127 & -2.1173 & -2.1914 & -1.9148 \\
	202   & 88    & 7.8803 & 0     & -2.3872 & -2.1527 & -1.9520 & -1.9762 & -1.9010 & -1.6470 \\
	203   & 88    & 7.7363 & 0     & -1.4437 & -1.7082 & -1.5077 & -1.3335 & -1.3730 & -1.3051 \\
	204   & 88    & 7.6366 & 0     & -1.2218 & -1.3973 & -1.1971 & -1.2360 & -1.1520 & -1.0045 \\
	205   & 89    & 8.0932 & 0     & -1.0969 & -2.4696 & -2.2674 & -2.0889 & -2.1312 & -1.8515 \\
	206   & 89    & 7.9583 & 0     & -1.6021 & -2.0673 & -1.8653 & -1.3750 & -1.3382 & -1.5490 \\
	207   & 89    & 7.8449 & 0     & -1.5086 & -1.7224 & -1.5206 & -1.3574 & -1.3750 & -1.2760 \\
	208   & 89    & 7.7286 & 0     & -1.0132 & -1.3595 & -1.1579 & -0.6785 & -0.5748 & -0.9364 \\
	208   & 90    & 8.2020 & 0     & -2.6198 & -2.4620 & -2.2584 & -2.2735 & -2.1985 & -1.8051 \\
	210   & 90    & 8.0690 & 0     & -1.7959 & -2.0839 & -1.8806 & -1.9085 & -1.8301 & -1.5345 \\
	211   & 90    & 7.9375 & 0     & -1.3188 & -1.6844 & -1.4812 & -1.3145 & -1.2960 & -1.2046 \\
	212   & 91    & 8.4108 & 0     & -2.2366 & -2.7663 & -2.5615 & -2.0529 & -2.0004 & -2.0309 \\
	213   & 91    & 8.3844 & 0     & -2.1308 & -2.7031 & -2.4985 & -2.3432 & -2.3619 & -1.9940 \\
	215   & 91    & 8.2361 & 0     & -1.8539 & -2.2864 & -2.0819 & -1.9417 & -1.9427 & -1.6860 \\
	216   & 91    & 8.0993 & 2     & -0.9788 & -1.6378 & -1.4265 & -0.7237 & -0.9561 & -0.9593 \\
	216   & 92    & 8.5306 & 0     & -2.1612 & -2.8024 & -2.5962 & -2.6148 & -2.5441 & -2.0390 \\
	218   & 92    & 8.7748 & 0     & -3.4510 & -3.5262 & -3.3208 & -3.3519 & -3.2938 & -2.6983 \\
	219   & 93    & 9.2075 & 0     & -3.2441 & -4.3515 & -4.1450 & -4.0041 & -4.0452 & -3.4962 \\
	223   & 93    & 9.6504 & 0     & -5.6021 & -5.5184 & -5.3139 & -5.2123 & -5.2794 & -5.3452 \\
	225   & 93    & 8.8182 & 0     & -2.1871 & -3.3793 & -3.1725 & -3.0745 & -3.0632 & -2.6083 \\
	228   & 94    & 7.9402 & 0     & 0.3222 & -0.3549 & -0.1446 & -0.2144 & -0.1754 & 0.2239 \\
	229   & 94    & 7.5980 & 2     & 2.2601 & 1.0589 & 1.2762 & 1.5247 & 1.5507 & 1.1252 \\
	230   & 94    & 7.1785 & 0     & 2.0212 & 2.3818 & 2.5922 & 2.4671 & 2.5020 & 2.3075 \\
	231   & 94    & 6.8386 & 0     & 3.5987 & 3.7556 & 3.9658 & 3.9517 & 4.2917 & 3.6338 \\
	234   & 96    & 7.3653 & 0     & 2.2846 & 2.4459 & 2.6601 & 2.5551 & 2.6082 & 2.4021 \\
	236   & 96    & 7.0670 & 0     & 3.3554 & 3.6116 & 3.8260 & 3.6812 & 3.7185 & 3.5425 \\
	237   & 98    & 8.2200 & 2     & 0.0580 & 0.3857 & 0.6102 & 0.9290 & 0.9302 & 0.8602 \\
	240   & 98    & 7.7110 & 0     & 1.6119 & 1.8946 & 2.1129 & 2.0218 & 2.0646 & 1.9214 \\
	242   & 99    & 8.1601 & 2     & 1.4945 & 0.9121 & 1.1390 & 2.0714 & 1.8378 & 1.4903 \\
	243   & 100   & 8.6892 & 1     & -0.5954 & -0.5957 & -0.3724 & 0.8667 & 0.0249 & -0.0970 \\
	247   & 101   & 8.7644 & 1     & 0.0755 & -0.5084 & -0.2831 & 0.2091 & 0.0054 & 0.0711 \\
	251   & 102   & 8.7517 & 0     & -0.0160 & -0.2224 & 0.0029 & 0.1319 & 0.4853 & -0.0580 \\
	254   & 103   & 8.8218 & 3     & 1.2237 & 0.3384 & 0.5804 & 1.4500 & 1.1793 & 0.9884 \\
	255   & 104   & 9.0555 & 1     & 0.4896 & -0.4050 & -0.1740 & 1.1434 & 0.3150 & 0.6843 \\
	256   & 104   & 8.9257 & 0     & 0.3282 & -0.0911 & 0.1380 & 0.1053 & 0.1402 & 0.1547 \\
	256   & 105   & 9.3361 & 2     & 0.3854 & -0.7522 & -0.5153 & 0.5411 & 0.2246 & 0.4760 \\
	259   & 106   & 9.7651 & 2     & -0.3958 & -1.6609 & -1.4233 & -0.9860 & -0.9367 & -0.4093 \\
	260   & 106   & 9.9006 & 0     & -1.7678 & -2.2570 & -2.0268 & -2.0209 & -1.9894 & -1.9146 \\
	261   & 106   & 9.7137 & 2     & -0.7292 & -1.5412 & -1.3030 & -0.8716 & -0.7890 & -0.2402 \\
	261   & 107   & 10.5002 & 3     & -1.8928 & -3.0653 & -2.8213 & -2.4141 & -2.5845 & -1.7629 \\
	265   & 108   & 10.4703 & 0     & -2.7077 & -3.1370 & -2.9048 & -2.6917 & -2.3532 & -2.5175 \\
	266   & 108   & 10.3457 & 0     & -2.4037 & -2.8320 & -2.5991 & -2.5868 & -2.5642 & -2.3619 \\
	267   & 110   & 11.7768 & 0     & -5.0000 & -5.5563 & -5.3268 & -5.0711 & -4.8241 & -4.7157 \\
	270   & 110   & 11.1170 & 0     & -3.6882 & -4.1269 & -3.8935 & -3.8621 & -3.8351 & -3.3006 \\
	286   & 114   & 10.3551 & 0     & -0.6569 & -1.1086 & -0.8609 & -0.8646 & -0.8471 & -0.9908 \\
	288   & 114   & 10.0765 & 0     & -0.1851 & -0.3534 & -0.1036 & -0.1337 & -0.1342 & 0.0531 \\
	290   & 116   & 10.9968 & 0     & -2.0458 & -2.1865 & -1.9378 & -1.9130 & -1.8826 & -1.9792 \\
	292   & 116   & 10.7912 & 0     & -1.7959 & -1.6805 & -1.4300 & -1.4259 & -1.4144 & -1.5953 \\
	294   & 118   & 11.8673 & 0     & -3.1549 & -3.6977 & -3.4498 & -3.4021 & -3.3665 & -3.0170 \\
\end{longtable}
Figure (\ref{figure1}) shows the plots of the $\log [T_{1/2}$(s)] values for the $69$ neutron-deficient nuclei obtained from using the various models. The experimental data are included for comparison.  It is observed from the plot that the predicted values agree with the values obtained from experiment. 

\begin{figure}[H]
	\centering
	\includegraphics[scale=0.7]{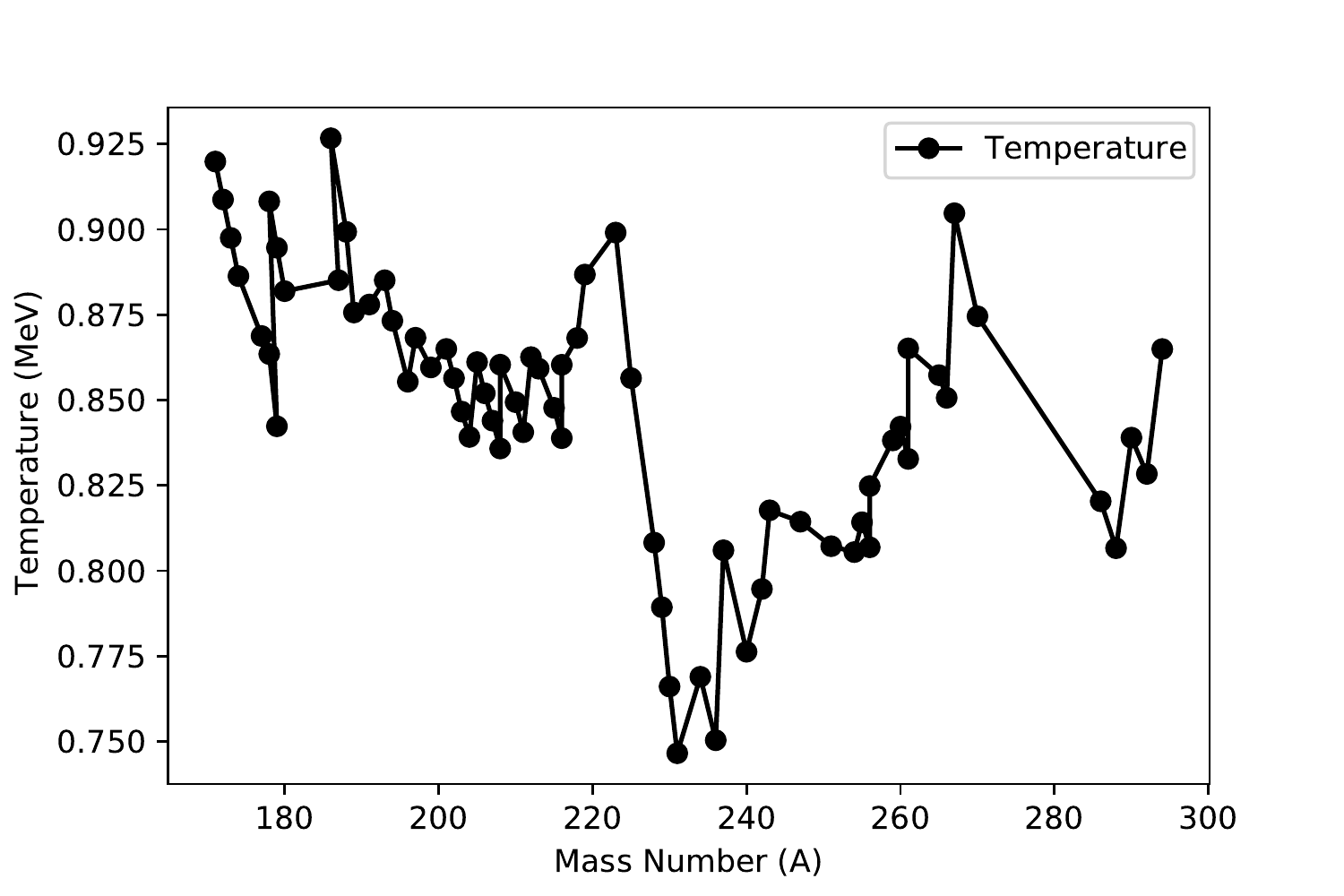} 
	\caption{Plot of the calculated temperature (MeV) values in the CPPMT for 69 neutron-deficient nuclei with respect to their mass number (A).}
	\label{figure2}
\end{figure}

In order to quantitatively evaluate the performance of the models, the root mean square error (RMSE) is calculated. The experimental $\alpha$-decay half-lives are retrieved from Refs. \cite{Kondev_2021,Wang2021}. Table \ref{table5} presents the computed root mean square error for all the models.  It can be observed that the ANN model gives the lowest RMSE, with a value of $0.3843$. The CPPMT (RMSE $ = 0.4963$) is found to give lower RMS error compared to the CPPM (RMSE $ = 0.5946$), indicating the importance of the use of temperature dependent potential. The NRB formula is also found to give lower RMSE value than the Royer formula. The calculated temperature values (in MeV) for the neutron-deficient nuclei in the CPPMT model are plotted with respect to the mass number (A) of these nuclei in Figure (\ref{figure2}).
\begin{table}[H]
	\centering
	\caption{The calculated RMSE values obtained for the neutron-deficient nuclei using CPPM, CPPMT, Royer, NRB, and ANN.}
	\begin{tabular}{lr}\hline
		\hline
		Models (Formulas) & $\sigma$ \\ \hline
		\hline
		CPPM & 0.5946\\ 
		CPPMT & 0.4963\\ 
		Royer & 0.4608\\ 
		NRB & 0.4413\\
		ANN & 0.3843 \\ \hline
	\end{tabular}
	\label{table5}
\end{table}

The energy released ($Q_\alpha$) during an $\alpha$-decay process is one of the important input parameters required to calculate $\alpha$-decay half-lives. To predict the half-lives of $\alpha$-decay from unmeasured neutron-deficient nuclei, the $Q_{\alpha}$ values are required. Previously the Weizs$\ddot{a}$cker-Skyrme-4 (WS4) and Weizs$\ddot{a}$cker-Skyrme-4+RBF (WS4+RBF) \cite{Wang2014} formulas were used to predict the $Q_{\alpha}$ values of unmeasured neutron-deficient nuclei \cite{Cui2019}. However in this work, we are motivated to use ANN to predict the $Q_\alpha$ values, following its success in predicting the $\alpha$-decay half-lives. To achieve this, an artificial neural network (ANN) is trained using $1021$ $Q_\alpha$ values of measured nuclei in the NUBASE2020 database. The $Q_\alpha$ values are again split into train (80 \% of data) and test (20 \% of data) sets. Procedure similar to that followed in the training of the half-lives is followed. After training and optimizations, the root mean square errors obtained on the train and test sets are given in Table \ref{table6}.
\begin{table}[H]
	\caption{The $\sigma$ values between the experimental and predicted  $Q_\alpha$ values for the training and test set.}
	\begin{center}
		\begin{tabular}{lr}\hline
			\hline
			Artificial Neural Network (ANN) & $\sigma$ \\ \hline
			\hline
			Train & 0.1684 \\ 
			Test & 0.1802 \\ 
			\hline
		\end{tabular}
	\end{center}
	\label{table6}
\end{table}

\begin{figure}[H]
	\centering
	\includegraphics[scale=0.7]{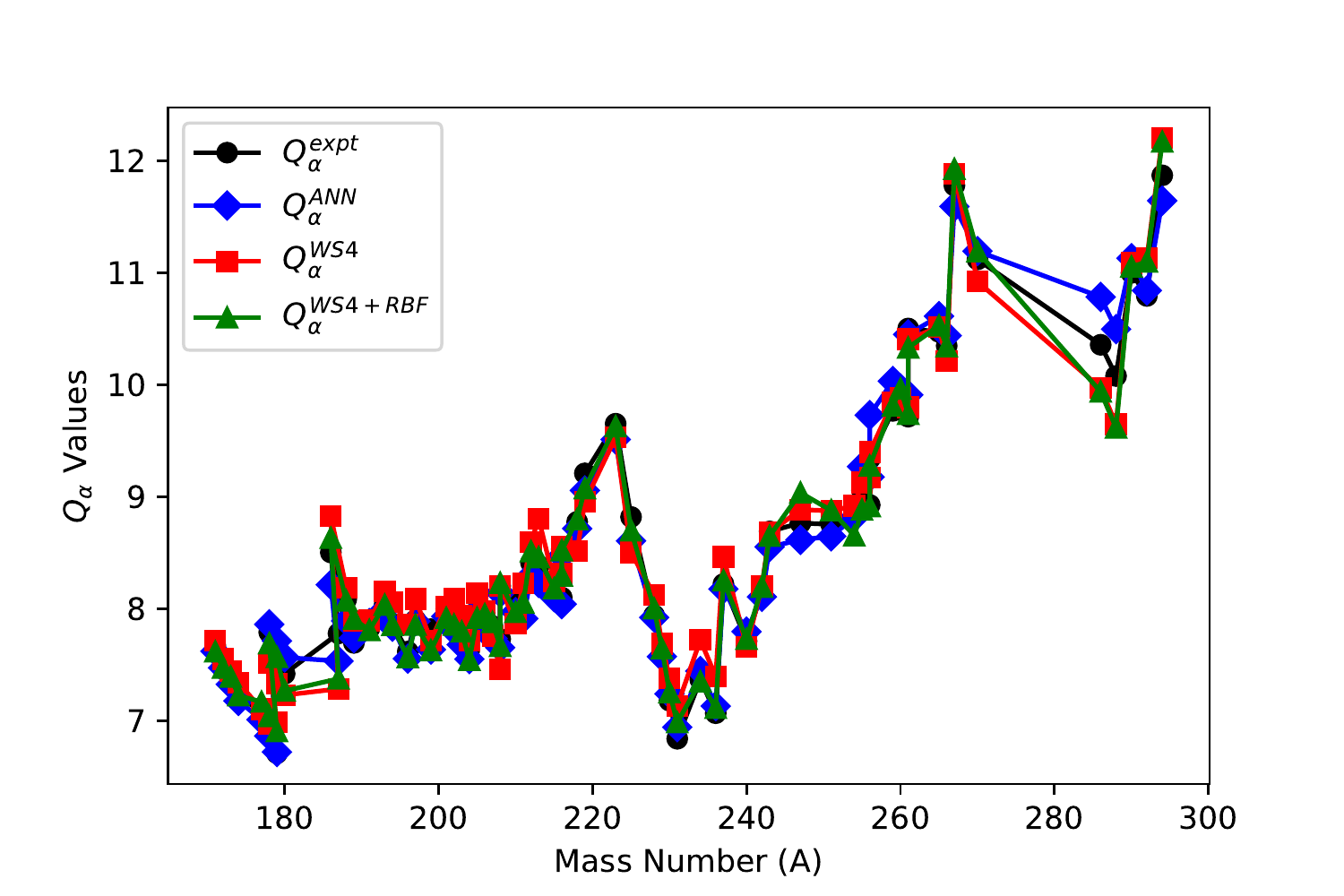}  
	\caption{ Plots of the experimental and predicted $Q_\alpha$ values for 69 neutron-deficient nuclei using WS4, WS4+RBF, and ANN models.}
	\label{figure3}
\end{figure}

In order to compare the performance of the ANN predictions of $Q_{\alpha}$ values with existing theories, we have used the ANN model to predict $Q_{\alpha}$ values for $69$ neutron-deficient nuclei. The outputs are compared with the predictions of WS4 and WS4+RBF models. Table \ref{table7} presents the root mean square error values obtained when the predictions of ANN, WS4, and WS4+RBF are compared with experimental values. With a standard deviation value of $0.1475$, the ANN model is found to give a slightly lower RMSE value than WS4+RBF theoretical model. The WS4+RBF, as expected, performs better than the WS4 formula. Figure (\ref{figure3}) shows the plots of the $Q_\alpha$ values predicted by WS4, WS4+RBF, and ANN models for the neutron-deficient nuclei.
\begin{table}[H]
	\caption{The computed root mean square errors ($\sigma$) obtained using WS4, WS4+RBF, and ANN models.}
	\begin{center}
		\begin{tabular}{lr}\hline
			\hline
			Models & $\sigma$ \\ \hline
			\hline
			WS4 & 0.2038\\
			WS4+RBF & 0.1565\\
			ANN &0.1475\\ \hline
		\end{tabular}
	\end{center}
	\label{table7}
\end{table}

\begin{figure}[H]
	\centering
	\includegraphics[scale=0.7]{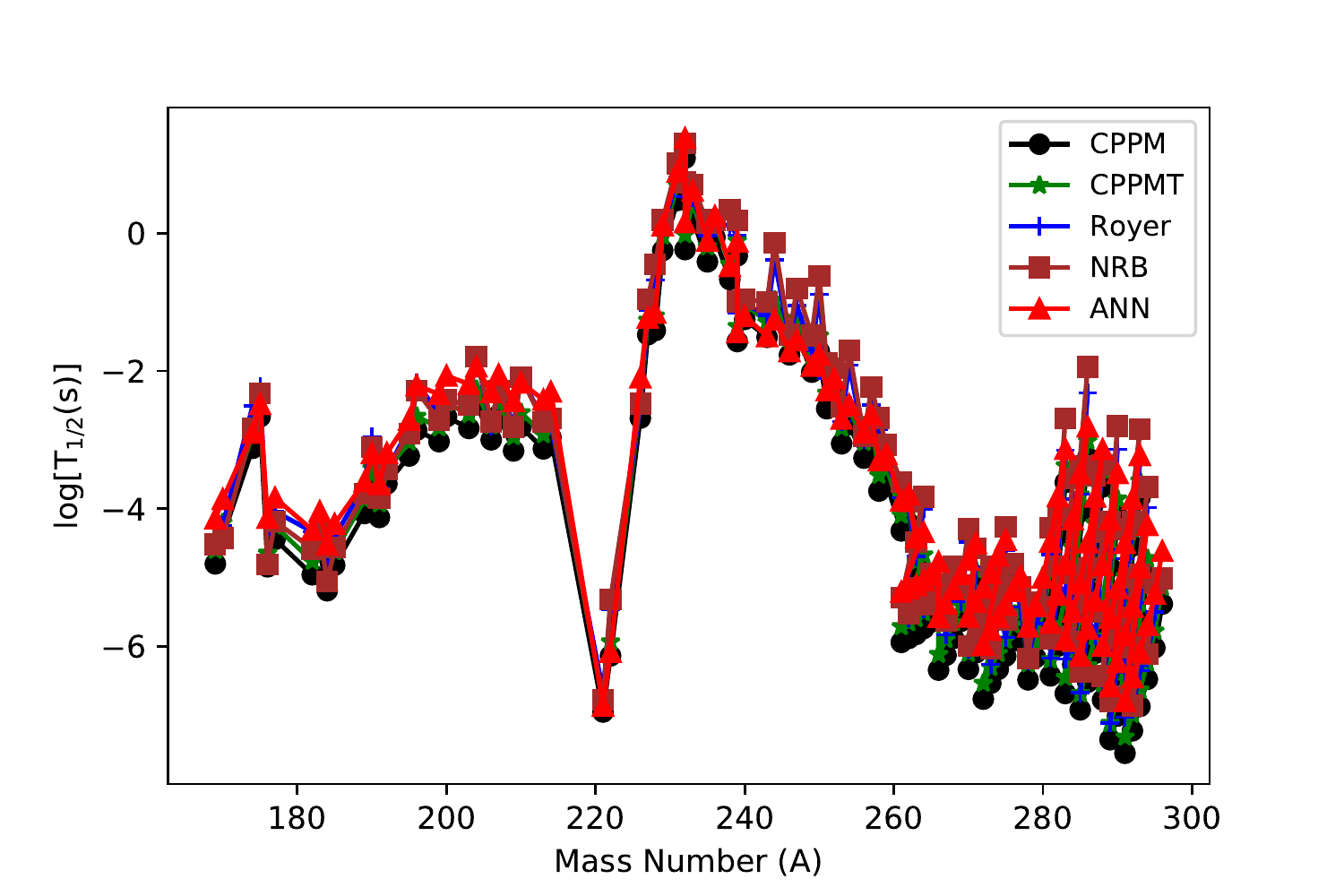} 
	\caption{ Plots of the predicted $\alpha$-decay half-lives for the neutron-deficient nuclei using the various models.}
	\label{figure4}
\end{figure}

Since the trained ANN model performed very well in predicting the $Q_{\alpha}$ values, we are now on track to predict the $\alpha$-decay half-lives of unmeasured neutron-deficient nuclei. The $Q_{\alpha}$ values predicted by the ANN model (denoted as $Q_{\alpha}^{ANN}$) will be used as part of the input values. The $\alpha$-decay half-lives of the neutron-deficient nuclei within the range of $80 \le Z \le 120$ and $169 \le A \le 296$ will then be predicted using CPPM, CPPMT, Royer, NRB, and the trained artificial neural network (denoted as $T^{ANN}$) models. The angular momentum $\ell$ carried by the emitted $\alpha$ particle has been taken to be zero for all nuclei. Table \ref{table9} presents the predicted half-lives for $\alpha$-decay of the 126 unmeasured neutron-deficient nuclei using the various theoretical models and $T^{ANN}$. The third to fifth columns of the Table show the $Q_{\alpha}$ values predicted by the WS4 ($Q_{\alpha}^{WS4}$), WS4+RBF ($Q_{\alpha}^{WS4+RBF}$), and ANN ($Q_{\alpha}^{ANN}$) models. The sixth to tenth columns show the predictions using CPPM, CPPMT, Royer, NRB, and $T^{ANN}$. The last column is the previous theoretical calculations of Cui et al. \cite{Cui2019} using generalized liquid drop model (GLDM). A careful observation of the values indicate that the results are close to that predicted earlier by Cui et al. \cite{Cui2019}. Figure (\ref{figure4}) shows the plots of the predicted log[T$_{1/2}$(s)] values using the various models.

\begin{small}
	\begin{longtable}{cccccccccccc}
		\caption{The predicted log[T$_{1/2}$(s)] values for 126 unmeasured neutron-deficient nuclei within the range of $80 \le Z \le 120$ using $Q_\alpha$ values predicted by ANN ($Q_{\alpha}^{ANN}$). Previous theoretical predictions by Cui et al. \cite{Cui2019} using GLDM are included for comparison. The $Q_{\alpha}^{WS4+RBF}$ values have been taken from \cite{Wang2014}}.\\ \hline
		\hline
		& & & & \multicolumn{8}{c} {log$[T_{1/2} (s)]$} \\ \cline{6 - 11}\\
		A & Z & $Q_{\alpha}^{WS4}$ & $Q_{\alpha}^{WS4+RBF}$ & $Q_{\alpha}^{ANN}$ & CPPM & CPPMT & Royer & NRB & $T^{ANN}$ & GLDM \\
		\hline
		\hline
		\endfirsthead
		\multicolumn{9}{l}
		{\tablename\ \thetable\ -- \textit{Contd}} \\ \hline
		\hline
		&&&&\multicolumn{8}{c}{log$[T_{1/2} (s)]$} \\ \cline{6 - 11}
		A & Z & $Q_{\alpha}^{WS4}$&$Q_{\alpha}^{WS4+RBF}$& $Q_{\alpha}^{ANN}$ & CPPM & CPPMT & Royer & NRB & ANN  & GLDM \\ \hline
		\hline
		\endhead
		\hline \multicolumn{11}{r}{\textit{Contd}} \\
		\endfoot
		\hline
		\endlastfoot
		\label{table9}
		
		169   & 80    & 7.8870 & 7.7235 & 7.9151 & -4.7990 & -4.6030 & -4.3998 & -4.5162 & -4.1667 & -4.2020 \\
		170   & 80    & 7.8490 & 7.7214 & 7.7667 & -4.3969 & -4.2017 & -4.2443 & -4.4235 & -3.8227 & -4.4034 \\
		174   & 81    & 7.5690 & 7.6406 & 7.4513 & -3.1224 & -2.9271 & -2.5043 & -2.8317 & -2.8110 & -2.9245 \\
		175   & 81    & 7.4680 & 7.5169 & 7.3018 & -2.6660 & -2.4715 & -2.2307 & -2.3293 & -2.3871 & -2.7747 \\
		176   & 82    & 7.8730 & 8.1468 & 8.1580 & -4.8419 & -4.6449 & -4.6765 & -4.8072 & -4.2548 & -3.6289 \\
		177   & 82    & 7.6880 & 7.8397 & 8.0084 & -4.4440 & -4.2477 & -4.0379 & -4.1846 & -3.9177 & -2.2000 \\
		182   & 83    & 8.3450 & 8.4631 & 8.2967 & -4.9568 & -4.7610 & -4.3330 & -4.5801 & -4.4132 & -3.1180 \\
		183   & 83    & 8.0580 & 8.2330 & 8.1659 & -4.6157 & -4.4205 & -4.2003 & -4.3560 & -4.1222 & -3.3507 \\
		184   & 84    & 9.1510 & 9.2195 & 8.5091 & -5.1886 & -4.9912 & -5.0178 & -5.0579 & -4.6249 & -6.6498 \\
		185   & 84    & 8.9660 & 8.8820 & 8.3642 & -4.8188 & -4.6219 & -4.4072 & -4.5643 & -4.2987 & -5.5258 \\
		189   & 85    & 8.4560 & 8.0567 & 8.2046 & -4.0686 & -3.8707 & -3.6550 & -3.7821 & -3.6884 & -4.1035 \\
		190   & 85    & 8.1530 & 7.9320 & 8.0399 & -3.6110 & -3.4135 & -2.9546 & -3.1026 & -3.2766 & -2.9747 \\
		191   & 86    & 8.4930 & 8.3920 & 8.3470 & -4.1255 & -3.9259 & -3.7059 & -3.8478 & -3.7540 & -3.9626 \\
		192   & 86    & 8.2250 & 8.1818 & 8.1697 & -3.6383 & -3.4390 & -3.4530 & -3.4299 & -3.2933 & -3.7959 \\
		195   & 87    & 8.3190 & 8.1319 & 8.1451 & -3.2338 & -3.0331 & -2.8230 & -2.9083 & -2.8545 & -3.2933 \\
		196   & 87    & 8.2120 & 7.8721 & 8.0140 & -2.8593 & -2.6590 & -2.1740 & -2.2908 & -2.4014 & -3.0205 \\
		199   & 88    & 8.1950 & 8.1145 & 8.1821 & -3.0260 & -2.8243 & -2.6138 & -2.7142 & -2.5563 & -2.7144 \\
		200   & 88    & 8.0320 & 8.0609 & 8.0561 & -2.6643 & -2.4629 & -2.4759 & -2.4154 & -2.1597 & -2.6271 \\
		203   & 89    & 8.5130 & 8.3122 & 8.2246 & -2.8343 & -2.6316 & -2.4408 & -2.4945 & -2.3111 & -3.4123 \\
		204   & 89    & 8.3610 & 8.0277 & 8.0985 & -2.4701 & -2.2676 & -1.7684 & -1.7958 & -1.9773 & -3.1361 \\
		206   & 90    & 8.5770 & 8.4303 & 8.3932 & -2.9993 & -2.7951 & -2.7990 & -2.7347 & -2.4549 & 2.3655 \\
		207   & 90    & 8.3660 & 8.2410 & 8.2671 & -2.6424 & -2.4384 & -2.2359 & -2.2877 & -2.1209 & -2.6925 \\
		209   & 91    & 8.5970 & 8.7452 & 8.5621 & -3.1601 & -2.9546 & -2.7734 & -2.8115 & -2.5994 & -3.1007 \\
		210   & 91    & 8.3590 & 8.5251 & 8.4361 & -2.8108 & -2.6054 & -2.0869 & -2.0940 & -2.2659 & -2.5498 \\
		213   & 92    & 8.8220 & 8.7026 & 8.6599 & -3.1307 & -2.9241 & -2.7112 & -2.7436 & -2.5706 & -3.4214 \\
		214   & 92    & 9.1570 & 8.6648 & 8.5962 & -2.9633 & -2.7568 & -2.7634 & -2.6916 & -2.4186 & -4.5214 \\
		221   & 93    & 10.5510 & 10.5627 & 10.2851 & -6.9486 & -6.7464 & -6.6515 & -6.7726 & -6.9742 & -6.7122 \\
		222   & 93    & 10.0770 & 9.9496 & 9.9162 & -6.1337 & -5.9301 & -5.4600 & -5.3107 & -6.1410 & -5.7932 \\
		226   & 94    & 8.7920 & 8.9110 & 8.6969 & -2.6852 & -2.4761 & -2.5152 & -2.4735 & -2.2448 & -2.8894 \\
		227   & 94    & 8.5180 & 8.5648 & 8.2903 & -1.4747 & -1.2649 & -1.1222 & -0.9567 & -1.0712 & -2.0070 \\
		228   & 95    & 8.7070 & 8.6537 & 8.3877 & -1.4136 & -1.2020 & -0.6793 & -0.4516 & -0.9432 & -2.3605 \\
		229   & 95    & 8.3210 & 8.1832 & 8.0231 & -0.2536 & -0.0417 & 0.0468 & 0.1892 & 0.1426 & -1.1561 \\
		231   & 96    & 8.1600 & 7.9503 & 7.9165 & 0.4742 & 0.6881 & 0.8114 & 1.0090 & 0.8291 & -0.3556 \\
		232   & 96    & 7.9950 & 7.7965 & 7.7383 & 1.0864 & 1.3004 & 1.2336 & 1.2970 & 1.3111 & 0.0607 \\
		232   & 97    & 8.6200 & 8.3928 & 8.2418 & -0.2396 & -0.0240 & 0.5248 & 0.7347 & 0.1942 & -1.7167 \\
		233   & 97    & 8.4670 & 8.2118 & 8.1018 & 0.2113 & 0.4270 & 0.5153 & 0.7029 & 0.5818 & -1.0482 \\
		235   & 98    & 8.8030 & 8.5756 & 8.4027 & -0.4168 & -0.1994 & -0.0400 & 0.1426 & -0.0364 & -1.8182 \\
		236   & 98    & 8.6370 & 8.4122 & 8.2898 & -0.0640 & 0.1535 & 0.1218 & 0.1889 & 0.2640 & -1.3645 \\
		238   & 99    & 8.8710 & 8.8323 & 8.5907 & -0.6741 & -0.4551 & 0.1134 & 0.3412 & -0.3543 & -1.9830 \\
		239   & 99    & 8.6660 & 8.6226 & 8.4779 & -0.3299 & -0.1107 & -0.0288 & 0.1800 & -0.0539 & -1.0434 \\
		239   & 100   & 9.3180 & 9.2752 & 9.0044 & -1.5734 & -1.3532 & -1.1562 & -0.9931 & -1.1934 & -2.6904 \\
		240   & 100   & 9.1130 & 9.0738 & 8.8916 & -1.2518 & -1.0313 & -1.0318 & -0.9633 & -0.9505 & -2.0964 \\
		243   & 101   & 9.1940 & 9.2682 & 9.0886 & -1.5116 & -1.2896 & -1.1940 & -0.9960 & -1.2614 & -1.8962 \\
		244   & 101   & 9.2850 & 9.3598 & 8.9778 & -1.1967 & -0.9744 & -0.3868 & -0.1428 & -1.0136 & -2.3125 \\
		246   & 102   & 10.0020 & 10.0677 & 9.2872 & -1.7673 & -1.5438 & -1.5352 & -1.4744 & -1.5546 & -3.9101 \\
		247   & 102   & 9.8410 & 9.9120 & 9.1764 & -1.4601 & -1.2362 & -1.0525 & -0.8081 & -1.3344 & -3.1068 \\
		249   & 103   & 9.8930 & 9.9942 & 9.4866 & -2.0175 & -1.7924 & -1.7061 & -1.4856 & -1.8338 & -2.8153 \\
		250   & 103   & 9.5980 & 9.6889 & 9.3756 & -1.7168 & -1.4914 & -0.8859 & -0.6261 & -1.5944 & -1.9626 \\
		251   & 104   & 9.8240 & 9.8773 & 9.7971 & -2.5478 & -2.3217 & -2.1050 & -1.8805 & -2.2510 & -2.2034 \\
		252   & 104   & 9.5560 & 9.5519 & 9.6860 & -2.2597 & -2.0332 & -2.0169 & -1.9635 & -2.0149 & -1.9706 \\
		253   & 105   & 9.8040 & 9.7829 & 10.1091 & -3.0566 & -2.8296 & -2.7369 & -2.5234 & -2.6619 & -1.6840 \\
		254   & 105   & 9.5950 & 9.5263 & 9.9977 & -2.7794 & -2.5520 & -1.9171 & -1.7028 & -2.4251 & -0.9586 \\
		256   & 106   & 9.7470 & 9.6551 & 10.3055 & -3.2651 & -3.0368 & -2.9989 & -2.9419 & -2.8264 & -1.7190 \\
		257   & 106   & 9.7110 & 9.6039 & 10.1769 & -2.9495 & -2.7206 & -2.4966 & -2.2387 & -2.5495 & -0.5361 \\
		258   & 107   & 10.2050 & 10.1217 & 10.6190 & -3.7425 & -3.5134 & -2.8536 & -2.6830 & -3.2131 & -1.5421 \\
		259   & 107   & 10.2430 & 10.1697 & 10.5628 & -3.6176 & -3.3883 & -3.3081 & -3.0738 & -3.1183 & -1.4237 \\
		261   & 108   & 10.9560 & 10.8824 & 10.9797 & -4.3210 & -4.0915 & -3.8401 & -3.6084 & -3.7525 & -2.8386 \\
		262   & 108   & 11.0170 & 10.9343 & 10.9286 & -4.2137 & -3.9840 & -3.9429 & -3.8976 & -3.6363 & -3.8570 \\
		263   & 109   & 11.7210 & 11.5862 & 11.4034 & -4.9987 & -4.7693 & -4.6934 & -4.4802 & -4.4301 & -4.2336 \\
		264   & 109   & 11.6690 & 11.5516 & 11.3524 & -4.8980 & -4.6682 & -4.0033 & -3.8277 & -4.3126 & -5.0141 \\
		261   & 110   & 12.1470 & 12.0874 & 11.9956 & -5.9378 & -5.7091 & -5.4160 & -5.2881 & -5.4567 & -5.1707 \\
		262   & 110   & 12.2240 & 12.1576 & 11.9611 & -5.8797 & -5.6510 & -5.5802 & -5.5186 & -5.3783 & -5.5287 \\
		263   & 110   & 12.3370 & 12.2642 & 11.9268 & -5.8216 & -5.5928 & -5.3114 & -5.1460 & -5.3003 & -5.0391 \\
		264   & 110   & 12.3870 & 12.3230 & 11.8801 & -5.7369 & -5.5079 & -5.4476 & -5.3933 & -5.1938 & -5.9788 \\
		265   & 110   & 12.3340 & 12.2357 & 11.8285 & -5.6413 & -5.4121 & -5.1430 & -4.9368 & -5.0755 & -5.1630 \\
		266   & 110   & 12.1720 & 12.1327 & 11.7170 & -5.4173 & -5.1875 & -5.1366 & -5.0907 & -4.8168 & -5.6840 \\
		266   & 111   & 12.7280 & 12.6966 & 12.3057 & -6.3402 & -6.1116 & -5.4305 & -5.3582 & -5.8294 & -6.7167 \\
		267   & 111   & 12.5450 & 12.5439 & 12.1922 & -6.1250 & -5.8958 & -5.8310 & -5.6302 & -5.5700 & -5.2807 \\
		268   & 111   & 12.2400 & 12.3038 & 12.0656 & -5.8794 & -5.6495 & -4.9710 & -4.8395 & -5.2807 & -5.7645 \\
		269   & 111   & 11.9250 & 12.0929 & 11.9402 & -5.6315 & -5.4008 & -5.3436 & -5.1125 & -4.9929 & -4.1785 \\
		270   & 111   & 11.6370 & 11.8923 & 11.8203 & -5.3906 & -5.1592 & -4.4851 & -4.2940 & -4.7148 & -4.4802 \\
		271   & 111   & 11.3730 & 11.5357 & 11.6883 & -5.1189 & -4.8867 & -4.8393 & -4.5764 & -4.4084 & -3.1158 \\
		270   & 112   & 12.2850 & 12.4256 & 12.4175 & -6.3279 & -6.0980 & -6.0412 & -5.9927 & -5.7475 & -5.2774 \\
		271   & 112   & 12.0610 & 12.2550 & 12.2921 & -6.0896 & -5.8589 & -5.5892 & -5.3513 & -5.4609 & -4.1574 \\
		272   & 112   & 11.8620 & 12.0421 & 12.1730 & -5.8598 & -5.6283 & -5.5780 & -5.5372 & -5.1887 & -4.5171 \\
		273   & 112   & 11.6400 & 11.8677 & 12.0494 & -5.6163 & -5.3841 & -5.1265 & -4.8383 & -4.9047 & -3.4559 \\
		274   & 112   & 11.5480 & 11.5164 & 11.9118 & -5.3387 & -5.1055 & -5.0631 & -5.0325 & -4.5852 & -4.0048 \\
		275   & 112   & 11.7410 & 11.7436 & 11.7868 & -5.0826 & -4.8486 & -4.6052 & -4.2638 & -4.2952 & -3.7825 \\
		272   & 113   & 12.5120 & 12.7064 & 12.7716 & -6.7611 & -6.5313 & -5.8437 & -5.7540 & -6.2190 & -5.8827 \\
		273   & 113   & 12.3250 & 12.4612 & 12.6459 & -6.5315 & -6.3009 & -6.2548 & -6.0278 & -5.9319 & -4.4584 \\
		274   & 113   & 12.0840 & 12.1525 & 12.5365 & -6.3296 & -6.0983 & -5.4128 & -5.2648 & -5.6819 & -5.0794 \\
		275   & 113   & 11.9340 & 12.0250 & 12.4338 & -6.1376 & -5.9056 & -5.8677 & -5.6133 & -5.4471 & -3.7878 \\
		276   & 113   & 12.0640 & 12.0436 & 12.3179 & -5.9161 & -5.6834 & -5.0025 & -4.7967 & -5.1825 & -4.9914 \\
		277   & 113   & 12.2010 & 12.1090 & 12.1994 & -5.6852 & -5.4516 & -5.4229 & -5.1381 & -4.9113 & -4.4283 \\
		278   & 114   & 12.5190 & 12.5039 & 12.7329 & -6.4816 & -6.2495 & -6.2041 & -6.1706 & -5.7936 & -5.3497 \\
		279   & 114   & 12.4300 & 12.3872 & 12.5629 & -6.1597 & -5.9263 & -5.6689 & -5.3498 & -5.4047 & -4.5918 \\
		280   & 114   & 12.2260 & 12.1816 & 12.3383 & -5.7192 & -5.4842 & -5.4419 & -5.4159 & -4.8911 & -4.8729 \\
		281   & 114   & 11.8160 & 11.7729 & 12.0572 & -5.1449 & -4.9080 & -4.6629 & -4.2787 & -4.2893 & -3.4660 \\
		282   & 114   & 11.3780 & 11.3363 & 11.7095 & -4.3998 & -4.1606 & -4.1222 & -4.0993 & -3.6364 & -3.1518 \\
		283   & 114   & 10.8790 & 10.8393 & 11.3618 & -3.6170 & -3.3755 & -3.1532 & -2.6909 & -3.0090 & -1.4342 \\
		281   & 115   & 12.2030 & 12.1640 & 12.8276 & -6.4237 & -6.1899 & -6.1691 & -5.8820 & -5.6719 & -3.8633 \\
		282   & 115   & 11.7770 & 11.7367 & 12.6022 & -5.9924 & -5.7570 & -5.0622 & -4.8337 & -5.1561 & -3.9393 \\
		283   & 115   & 11.3240 & 11.2842 & 12.3640 & -5.5215 & -5.2844 & -5.2660 & -4.9299 & -4.6075 & -2.0953 \\
		284   & 115   & 10.9330 & 10.8941 & 12.0163 & -4.8003 & -4.5608 & -3.8615 & -3.5616 & -3.9542 & -1.9666 \\
		285   & 115   & 10.7300 & 10.6920 & 11.6687 & -4.0433 & -3.8014 & -3.7881 & -3.3798 & -3.3014 & -0.7282 \\
		286   & 115   & 10.5010 & 10.4646 & 11.3257 & -3.2589 & -3.0148 & -2.3179 & -1.9406 & -2.7463 & -0.7932 \\
		283   & 116   & 12.1070 & 12.0713 & 13.0928 & -6.6803 & -6.4461 & -6.1769 & -5.8667 & -5.9399 & -3.4248 \\
		284   & 116   & 11.8320 & 11.7950 & 12.8660 & -6.2570 & -6.0213 & -5.9681 & -5.9302 & -5.4211 & -3.4535 \\
		285   & 116   & 11.5490 & 11.5115 & 12.6434 & -5.8290 & -5.5916 & -5.3313 & -4.9603 & -4.9118 & -2.3188 \\
		286   & 116   & 11.3120 & 11.2755 & 12.3235 & -5.1861 & -4.9464 & -4.8944 & -4.8603 & -4.2727 & -2.3585 \\
		287   & 116   & 11.2840 & 11.2478 & 11.9755 & -4.4526 & -4.2105 & -3.9658 & -3.5206 & -3.6192 & -1.7959 \\
		288   & 116   & 11.2900 & 11.2551 & 11.6278 & -3.6833 & -3.4388 & -3.3945 & -3.3610 & -3.0211 & -2.4168 \\
		285   & 117   & 12.4450 & 12.4112 & 13.3513 & -6.9169 & -6.6824 & -6.6684 & -6.3692 & -6.1928 & -3.8153 \\
		286   & 117   & 12.2670 & 12.2324 & 13.1308 & -6.5154 & -6.2793 & -5.5673 & -5.3732 & -5.6882 & -4.5986 \\
		287   & 117   & 12.0520 & 12.0169 & 12.9076 & -6.0969 & -5.8591 & -5.8452 & -5.4999 & -5.1777 & -3.1215 \\
		288   & 117   & 11.9820 & 11.9472 & 12.6317 & -5.5592 & -5.3193 & -4.6020 & -4.3407 & -4.5929 & -3.9872 \\
		289   & 117   & 11.9870 & 11.9523 & 12.2830 & -4.8475 & -4.6051 & -4.5925 & -4.1833 & -3.9383 & -3.1079 \\
		290   & 117   & 11.8390 & 11.8056 & 11.9347 & -4.1017 & -3.8569 & -3.1358 & -2.7982 & -3.3030 & -3.6326 \\
		288   & 118   & 12.6160 & 12.5833 & 13.3997 & -6.7737 & -6.5373 & -6.4756 & -6.4262 & -5.9648 & -4.4597 \\
		289   & 118   & 12.5920 & 12.5590 & 13.1729 & -6.3587 & -6.1206 & -5.8473 & -5.4854 & -5.4461 & -3.9586 \\
		290   & 118   & 12.6010 & 12.5675 & 12.9401 & -5.9191 & -5.6792 & -5.6172 & -5.5723 & -4.9134 & -4.5258 \\
		291   & 118   & 12.4200 & 12.3873 & 12.5911 & -5.2286 & -4.9862 & -4.7235 & -4.2933 & -4.2585 & -3.7190 \\
		292   & 118   & 12.2400 & 12.2080 & 12.2574 & -4.5377 & -4.2928 & -4.2324 & -4.1871 & -3.6322 & -3.8794 \\
		293   & 118   & 12.2420 & 12.2100 & 11.9355 & -3.8411 & -3.5939 & -3.3518 & -2.8461 & -3.1053 & -3.4522 \\
		289   & 119   & 13.1750 & 13.1266 & 13.8570 & -7.3501 & -7.1149 & -7.1084 & -6.7948 & -6.6725 & -4.6655 \\
		290   & 119   & 13.0670 & 13.0351 & 13.6591 & -7.0071 & -6.7703 & -6.0430 & -5.8820 & -6.2197 & -5.7799 \\
		291   & 119   & 13.0480 & 13.0161 & 13.4411 & -6.6179 & -6.3795 & -6.3717 & -6.0151 & -5.7210 & -4.5376 \\
		292   & 119   & 12.9020 & 12.8696 & 13.2155 & -6.2030 & -5.9629 & -5.2294 & -5.0040 & -5.2050 & -5.4248 \\
		293   & 119   & 12.7150 & 12.6834 & 12.9101 & -5.6173 & -5.3748 & -5.3666 & -4.9556 & -4.5994 & -4.0057 \\
		294   & 119   & 12.7260 & 12.6947 & 12.5855 & -4.9678 & -4.7228 & -3.9822 & -3.6844 & -3.9907 & -5.0985 \\
		291   & 120   & 13.5090 & 13.4787 & 14.1035 & -7.5470 & -7.3114 & -7.0234 & -6.7247 & -6.8978 & -4.9872 \\
		292   & 120   & 13.4680 & 13.4370 & 13.9114 & -7.2213 & -6.9842 & -6.9149 & -6.8539 & -6.4583 & -5.4271 \\
		293   & 120   & 13.4000 & 13.3695 & 13.7083 & -6.8678 & -6.6291 & -6.3453 & -5.9911 & -5.9938 & -4.8996 \\
		294   & 120   & 13.2420 & 13.2109 & 13.4888 & -6.4741 & -6.2337 & -6.1627 & -6.1055 & -5.4916 & -5.1720 \\
		295   & 120   & 13.2720 & 13.2410 & 13.2420 & -6.0166 & -5.7743 & -5.4978 & -5.0826 & -4.9614 & -4.8297 \\
		296   & 120   & 13.3430 & 13.3124 & 12.9143 & -5.3822 & -5.1373 & -5.0653 & -5.0094 & -4.3504 & -5.4609 \\
	\end{longtable}
\end{small}

\section{Conclusion}
\label{conclusion}	

In this study, $\alpha$-decay half-lives of some neutron-deficient nuclei within the range of $ 80\le Z \le 118$  have been calculated using Coulomb and proximity potential model (CPPM), temperature dependent Coulomb and proximity potential model (CPPMT), Royer formula, New RenB (NRB) formula, and a trained artificial neural network ($T^{ANN}$) model. New coefficients were obtained for the Royer and NRB empirical formulas with the aid of a least square fit scheme and input data from NUBASE2020 database. When compared with the experimental data, all models are found to give very good predictions of the half-lives. The CPPMT was found to perform better than CPPM, indicating the importance of using temperature-dependent nuclear potentials. With a root mean square error of $0.3843$, the $T^{ANN}$ model is found to give the best performance in predicting the half-lives of the neutron-deficient nuclei. The second stage of the study was to predict the half-lives of $\alpha$-decay from unmeasured neutron-deficient nuclei. To achieve this, the $Q_{\alpha}$ values were required as inputs. Following the success of the ANN in predicting the half-lives, we were motivated to train another ANN to predict $Q_{\alpha}$ values (denoted as $Q_{\alpha}^{ANN}$). When compared to experimental $Q_{\alpha}$ values and theoretically predicted ones by WS4 and WS4+RBF formulas, the ANN model is found to give very good descriptions of the $Q_{\alpha}$ values. The $Q_{\alpha}^{ANN}$ values were then used as inputs to predict the half-lives of $\alpha$-decay from unmeasured neutron-deficient nuclei using CPPM, CPPMT, improved Royer formula, improved NRB formula and the $T^{ANN}$ model. The results of the predicted half-lives by our models are found to be in good agreements with those predicted using generalized liquid drop model (GLDM). This study concludes that half-lives of $\alpha$-decay from neutron-deficient nuclei can successfully be predicted using ANN, and this can contribute to the determination of nuclei at the driplines.


\bibliographystyle{unsrt}
\bibliography{Referencesalpha.bib}	

\end{document}